\documentclass[12pt,a4paper,final]{iopart}

\usepackage{iopams}
\usepackage{graphicx}
\usepackage[breaklinks=true,colorlinks=true,linkcolor=blue,urlcolor=blue,citecolor=blue]{hyperref}

\begin{document}
\title[Nanotechnology(2015)]{Straintronic spin-neuron}

\author{Ayan K. Biswas$^1$, Jayasimha Atulasimha$^2$, and Supriyo Bandyopadhyay$^1$}
\address{$^1$Department of Electrical and Computer Engineering, Virginia Commonwealth University, Richmond, Virginia 23284, USA}
\address{$^2$Department of Mechanical and Nuclear Engineering, Virginia Commonwealth University, Richmond, Virginia 23284, USA}
\eads{\mailto{sbandy@vcu.edu}}

\begin{abstract}
In artificial neural networks, neurons are usually implemented with highly dissipative CMOS-based
 operational
amplifiers. A more energy-efficient implementation is a ``spin-neuron'' realized with a
magneto-tunneling
junction (MTJ) that is switched with a spin-polarized current (representing weighted sum of
input currents) that either delivers
a spin transfer torque or induces domain wall motion in the soft layer of the MTJ.
Here, we propose and analyze a different type of spin-neuron in which the soft layer of the MTJ is switched
 with mechanical strain generated
by a voltage (representing weighted sum of input voltages) and term it
{\it  straintronic spin-neuron}.
It dissipates orders of magnitude less energy in threshold operations
than the traditional current-driven spin neuron at 0 K temperature and may even be faster.
We have also studied the room-temperature
firing behaviors of both types of spin neurons and find that thermal noise degrades the
performance of both types, but the current-driven type is degraded
 much more than
 the straintronic type if both are optimized for maximum energy-efficiency. On the other hand, if
 both are designed to have the same level of thermal degradation, then the current-driven version will
 dissipate orders of magnitude more energy than the straintronic version. Thus, the straintronic spin-neuron
 is superior to current-driven spin neurons.

\end{abstract}


\section{Introduction}

The building blocks of neural computing architectures are `neurons' usually connected to each other
and to external stimuli through programmable `synapses'.
The transfer function of a neuron can be expressed as
\begin{equation}
 O = f \left( \sum_i w_i x_i + b \right)
\end{equation}
where $f$ is some nonlinear function, $w_i$-s are programmable weights of synapses, $x_i$-s are the input signals
(representing dendrites),
$b$ is a fixed bias and $O$ is the output (representing a neuron's axon). In threshold operations, $f$ mimics a
Heaviside unit step function whose value is 1 if the
argument is positive and 0 otherwise.

Neural networks have myriad
topologies, such as cellular neural network \cite{Chua1988}, feed-forward network \cite{Ueda2011}, convolutional neural
network \cite{Simard2003}, hierarchical temporal memory
\cite{George2009}, etc. However, the basic unit of computing i.e., the neuron, remains more or less invariant
 across all topologies and its operation is governed by Equation (1). In a conventional
neural network, CMOS operational amplifiers carry out the threshold operation of Equation (1) \cite{Lippmann1987}
and dissipate exorbitant amounts of energy. To a large extent, this has stymied the progress of neural computing.
Alternate implementations to lower the  energy dissipation have been proposed in recent years
\cite{Datta2012,KaushikRoy2012,KaushikRoy2013} and utilize a magneto-tunneling junction (MTJ) whose
soft layer is an anisotropic nanomagnet with two stable magnetization
orientations.
Input variables are encoded in spin-polarized currents that are summed with variable weights to produce
a net spin polarized current which is driven through the nanomagnet. When the net current exceeds a
threshold value,  the magnetization of the soft layer rotates from one stable orientation to
the other, thereby changing the resistance of the MTJ abruptly.
This implements the threshold firing behavior of a neuron.
These types of artificial neurons have been termed `spin-neurons' and unlike CMOS-based neurons, they are `non-volatile' since
the final state of the neuron can be stored in the magnetization state of the nanomagnet
(and therefore the resistance of the MTJ) after the device is
powered down. In Ref. \cite{KaushikRoy2013}, the soft layer of the MTJ is a
nanomagnet possessing perpendicular magnetic anisotropy (see Figure 3(a) of Ref. \cite{KaushikRoy2013})
and it is switched with spin-polarized current generated via the giant
spin Hall effect \cite{Liu2012} which induces domain wall motion.
 This type of spin neurons belongs to the general class of (spin-polarized) {\it current driven} artificial neurons.

In this paper we propose and analyze a different type of spin-neuron implemented with MTJs having soft layers that are
magnetostrictive or multiferroic nanomagnets and
whose magnetizations are flipped with mechanical stress/strain generated by a voltage. We call them
`straintronic spin-neurons' and they are
{\it voltage-driven} as opposed to current-driven.
This has the advantage of further reducing the energy dissipation during firing.
Switching of multiferroic nanomagnets with voltage-generated stress has been proposed and/or demonstrated
by many groups
\cite{Ramesh2007,Brintlinger2010,Atulasimha2010,Buzzi2013,DSouzaArxiv} and is particularly useful
 for writing bits in non-volatile memory \cite{Tiercelin2011,Pertsev2009,Roy2013,Biswas2014a,Biswas2014b,Wang2014}. It can be also harnessed
for logic applications \cite{Atulasimha2010,Mohammed2011,Mohammed2012,Biswas2014c} and results in exceptionally
low dissipation.
Here, we propose it for neural applications. We compare the energy-efficiency of
a straintronic spin neuron with that of a traditional current-driven spin neuron and show that the former is
more
energy efficient. Finally, since magnetization dynamics is vulnerable to
thermal noise, we study the operation of spin-neurons at room temperature in the presence of thermal noise and compare
that with 0 K operation to assess the degree of thermal degradation. As expected, thermal noise has a deleterious effect on
the threshold behavior and seriously
degrades the abruptness of the firing action. The degradation
is far worse for the current-driven type than for the straintronic type.

\section{Straintronic Spin Neuron}

Figure 1 shows the schematic of a straintronic spin-neuron with programmable synapses. Inputs $x_i$-s and the
bias $b$ are in the form of voltages $V_i$-s and $b$, the latter being realized with a constant current source $I$
[$b = I \left(R_1 \parallel R_2 \parallel r_1 \parallel r_2 \parallel \cdot \cdot \cdot \parallel r_{N-1} \parallel r_N \right )$]. The voltage appearing at node $P$ is dropped
across the piezoelectric layer underneath the (shorted) contact pads A and A$^{\prime}$. It is a weighted
sum of input voltages and bias, and  is given by
\begin{equation}
V_P = \sum_{i=1}^N w_iV_i + b ,
\end{equation}
where $w_i = {{R_1 \parallel R_2 \parallel r_1 \parallel r_2 \parallel \cdot \cdot \cdot \parallel r_{i-1} \parallel r_{i+1}
\parallel \cdot \cdot \cdot \parallel r_N}\over{R_1 \parallel R_2 \parallel r_1 \parallel r_2 \parallel \cdot \cdot \cdot \parallel r_{i-1} \parallel r_{i+1}
\parallel \cdot \cdot \cdot \parallel r_N + r_i}}$. The resistances $R_1, R_2$ are the resistances of the piezoelectric
layer underneath the contact pads and $r_i$-s are the series resistances (connected to the input terminals) that implement the programmable weights. Equation
(2) is obtained from voltage superposition.

The magneto-tunneling junction (MTJ) in Figure 1 is the central unit of the neuron. It has a hard nanomagnet, a spacer layer and a soft magnetostrictive nanomagnet in contact
with the piezoelectric. All nanomagnets are shaped like elliptical disks. A bias magnetic field $B$ in the plane of the soft
nanomagnet directed along its minor axis makes the magnetization orientation of the soft nanomagnet bistable, with the two stable
directions shown as $\Psi_{\parallel}$ and $\Psi_{\perp}$ which subtend an angle of $\sim$90$^{\circ}$ between them
\cite{Tiercelin2011}.
The
hard nanomagnet is implemented with a synthetic anti-ferromagnet and its two stable magnetization
orientations are roughly along its major axis because of the extremely high shape
anisotropy that this nanomagnet possesses. The hard nanomagnet is placed such that
its major axis is collinear with one of the stable magnetization orientations of the soft nanomagnet (say $\Psi_{\parallel}$),
resulting in a ``skewed MTJ stack'' where the major axes of the two nanomagnets are at an angle. The
 hard nanomagnet is then magnetized permanently in the direction that is {\it anti-parallel} to $\Psi_{\parallel}$.
 Thus, when the soft nanomagnet is
 in the stable state $\Psi_{\parallel}$, the magnetizations of the hard and
soft layers of the MTJ are mutually anti-parallel, resulting in high MTJ resistance, while when the soft nanomagnet is in the other stable state
$\Psi_{\perp}$,
the magnetizations of the two layers are roughly perpendicular to each other, resulting in lower MTJ resistance. The ratio of the
high-to-low MTJ resistances is approximately $1/ \left (1 - \eta_1 \eta_2 \right )$, where the $\eta$-s are the spin injection/detection
efficiencies at the interfaces of the spacer with the two nanomagnets. We assume that at room temperature, $\eta_1 = \eta_2 \approx 70\%$ \cite{Salis2005} and therefore the resistance ratio will be
roughly 2:1. Higher resistance ratios exceeding 6:1 have been demonstrated at room temperature \cite{Ikeda2008},
but we will be conservative and assume the ratio to be 2:1.

\begin{figure}[!ht]%
 \centering
  \includegraphics[width=3.4in]{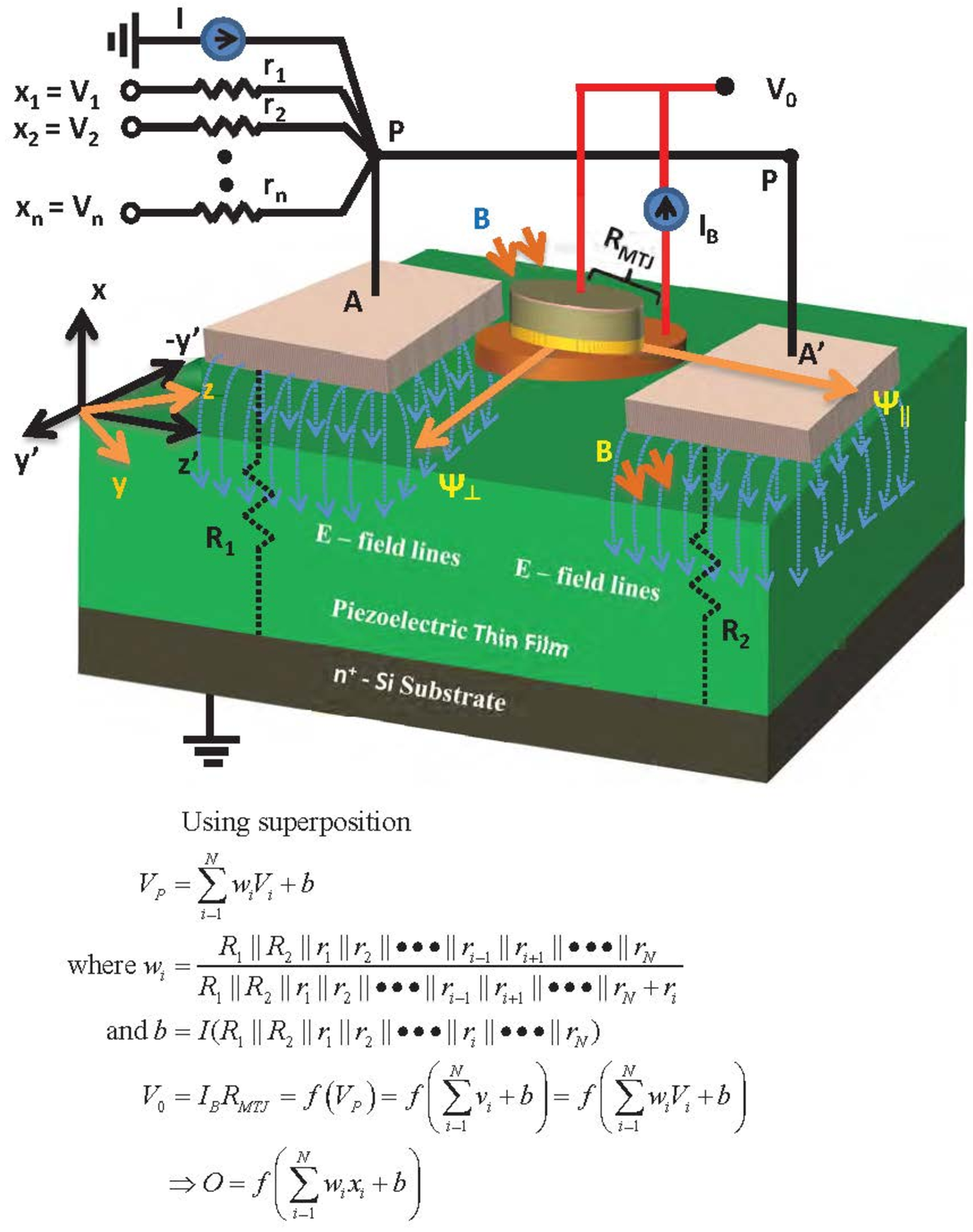}
 \caption[Caption of figure]{ Schematic of a straintronic spin-neuron implementing a step transfer function.
 The artificial synapses are realized with the passive resistors
$r_1 \cdots r_n$.}
  \label{fig:MTJ_SN}
\end{figure}

The electrodes A and A$^{\prime}$ are placed on the piezoelectric layer such that
the  line joining their centers is parallel to $\Psi_{\parallel}$ and hence also to the
major axis of the hard nanomagnet. Their lateral dimensions are of the same order as
and the inter-electrode separation is 1-2 times
the PZT thin film thickness.
When voltages are applied between the electrode pair  and the (grounded) conducting substrate,
electric fields are generated underneath
 the electrode pads in the PZT layer as shown in Figure 1. They produce out-of-plane
compressive
strain and in-plane tensile strain or vice versa, depending
on the polarity of the voltage in the PZT layer below the electrodes
\cite{Lynch2013}. These  strain fields interact and produce biaxial strain
between the electrodes (tensile along the line joining the electrodes and compressive in the perpendicular direction,
or vice versa) \cite{Lynch2013},
which is almost completely
transferred  to the magnetostrictive soft magnet since the latter's thickness is much smaller than that of
the strained PZT layer.

If the magnetostriction
coefficient of the soft magnet material is positive (e.g. Terfenol-D), then sufficient compressive stress resulting along
the line joining the electrode centers, i.e. in the
direction of $\Psi_{\parallel}$, will rotate the soft layer's magnetization to  $\Psi_{\perp}$, while sufficient tensile
stress  will keep the magnetization aligned along $\Psi_{\parallel}$.
The situation will be opposite if the magnetostriction coefficient of the soft magnet is negative (e.g. cobalt).
Since, for either sign of the magnetostriction coefficient, the sign of the stress (compressive or tensile) depends
on the polarity of the voltage applied between the electrodes and the grounded substrate, the magnetization of
the soft magnet can be aligned along either of the two stable orientations $\Psi_{\parallel}$ and $\Psi_{\perp}$ at will
by
simply choosing the voltage polarities (of the inputs and bias voltages).

There is a {\it minimum} stress (compressive/tensile) required to switch the magnetization of the soft nanomagnet of the MTJ
from one orientation (say, $\Psi_{\parallel}$)
to the other (say, $\Psi_{\perp}$) because the two stable states
are separated by an energy barrier \cite{Tiercelin2011, Biswas2014a, Biswas2014b, Biswas2014c} that needs to be overcome by
stress to make the switching occur. At 0 K temperature, this feature gives rise to a
sharp threshold in the switching behavior and makes it possible to mimic the sudden firing behavior of a
neuron. The energy barrier (and hence the threshold stress) depends on the permanent magnetic field $B$ and the size and shape of
the soft nanomagnet, if we ignore dipole coupling with the hard nanomagnet and any magneto-crystalline anisotropy (the
magnets are assumed to be amorphous). These parameters determine the
effective in-plane energy barrier between the two stable magnetization states $\Psi_{\parallel}$ and
$\Psi_{\perp}$ that must be overcome by stress
to switch the magnetization of the soft nanomagnet from one state to the other and therefore the MTJ resistance. The minimum stress needed for
switching (also called the `critical stress') at 0 K can be
found  by equating the stress anisotropy energy to the effective in-plane energy barrier.

The critical stress gives rise to a critical voltage $V_c$ for switching at 0 K. When the total
voltage $V_P$, appearing at node $P$, due to all weighted inputs and the bias, exceeds the critical voltage, the MTJ resistance
switches abruptly because the soft layer's magnetization rotates from one stable orientation to the other. If we bias the MTJ with a constant current source $I_B$ as shown in Fig. 1, then
the voltage across the MTJ is $V_{MTJ} = I_B R_{MTJ}$ where $R_{MTJ}$ is the MTJ resistance that
has a non-linear dependence on $V_P$ (abrupt switching when $V_P$ exceeds $V_c$). Because there is virtually
no electric field in the PZT layer directly underneath the magnet, we can ignore any potential drop
in this region and write the output voltage in Fig. 1 as $V_0 \approx V_{MTJ} = I_B R_{MTJ}$
which has a non-linear dependence on $V_P$ and hence exhibits a threshold behavior. Using
Equation (2), we can now write
\begin{equation}
V_0 = f \left ( V_P \right ) = f \left ( \sum_{i=1}^N w_iV_i + b \right ) ,
\end{equation}
which replicates the neural behavior.

\section{Simulation of neuron firing with the Landau-Lifshitz-Gilbert (LLG) equation}

We will design our straintronic spin-neuron by choosing the dimensions of the nanomagnet and the magnitude of the
permanent magnetic field to select the critical voltage $V_c$, and therefore the firing threshold.
We choose a soft nanomagnet of major axis  $a$ = 100 nm,  minor axis  $b$ = 42 nm and thickness $d$ = 16.5 nm, which ensures
that it has a single ferromagnetic domain \cite{Cowburn1999}. When stress is applied on the soft nanomagnet
by applying the voltage $V_P$ at the electrodes A and A$^{\prime}$,
which is the weighted sum of the inputs and bias, the magnetization vector of the soft nanomagnet experiences a torque that makes it rotate
and ultimately switch the MTJ resistance. The torque depends on the shape anisotropy energy of the soft nanomagnet,
the permanent magnetic field and the stress.

The shape
anisotropy energy  is a function of the
time-dependent magnetization orientation during switching and can be written as
\begin{eqnarray}
E_{sh}(t) & = & E_{s1}(t) {\sin}^2{\theta^{\prime}(t)} + E_{s2}(t) {\sin 2\theta^{\prime}(t)} +  \frac{\mu_{0}}{4} \Omega {M}^2_{s} ( N_{d-yy}+ N_{d-zz} )\nonumber\\
E_{s1}(t) & = & \left( \frac{\mu_{0}}{4} \right) \Omega {M}^2_{s} \left(2 N_{d-xx} - N_{d-yy} - N_{d-zz}\right){\cos}^2{\phi^{\prime}(t)} \nonumber\\
E_{s2}(t) & = & \left( \frac{\mu_{0}}{4} \right) \Omega {M}^2_{s} \left( N_{d-zz} -  N_{d-yy} \right){\sin \phi^{\prime}(t)},
\end{eqnarray}
where $\theta^{\prime}(t)$ and $\phi^{\prime}(t)$ are, respectively, the instantaneous polar and azimuthal angles of the
soft nanomagnet's magnetization vector in the primed reference frame shown in Figure 1. The unprimed reference frame is
such that the z-axis coincides with the soft nanomagnet's easy axis and y-axis with the hard axis. The primed reference
frame is obtained by rotating anticlockwise about the x-axis by 45$^\circ$.
The quantity
$M_s$ is the saturation magnetization of the soft nanomagnet, $N_{d-xx}$, $N_{d-yy}$ and $N_{d-zz}$
are the demagnetization factors
that can be evaluated from the nanomagnet's dimensions \cite{Beleggia2005}, $\mu_0$ is the permeability of free space,
and $\Omega = (\pi/4)abd$ is the soft nanomagnet's volume.

The permanent magnetic field contributes an additional term to the potential energy of the soft nanomagnet
given by
\begin{equation}
E_{m}(t) = \frac{1}{\sqrt{2}} M_s \Omega B \left [ {\cos \theta^{\prime}(t)}  - {\sin \theta^{\prime}(t)} {\sin \phi^{\prime}(t)}  \right ].
\end{equation}

The locations of the total potential energy $\left ( E_{sh} (\theta (t), \phi(t)) + E_m (\theta(t), \phi(t)) \right )$
minima (in the absence of stress) determine the stable magnetization
orientations of the soft nanomagnet and are found by
minimizing the total potential energy with respect to $\theta(t), \phi(t)$.
In our case, the
two stable orientations (the two degenerate energy minima)
turn out to be $\Psi_{\parallel}$ ($\theta = \theta_1 = 46.9^\circ, \phi = \phi_1 = 90^\circ$)
and $\Psi_{\perp}$
($\theta = \theta_0 = 133^\circ, \phi = \phi_0 = 90^\circ = \phi_1$) with an angular separation $\gamma$ of 86.1$^\circ$ (Figure 1)
between them when B is 0.14 T.
The in-plane energy barrier separating these two energy minima is 73.1 kT at room temperature
resulting in a tiny probability ($e^{-73.1} \sim 10^{-32}$)
of the neuron firing spontaneously at room temperature by switching between the stable states $\Psi_{\parallel}$ and
$\Psi_{\perp}$ per attempt. With an attempt frequency of 10$^{15}$ Hz, the mean time between successive spontaneous firing
(switching of the MTJ) would then be $10^{-15} \times 10^{32}$ = 10$^{17}$ seconds = 3$\times$10$^7$ centuries.

When voltages are applied at the electrodes A and A$^{\prime}$, the resulting stress
produced in the soft nanomagnet contributes a stress anisotropy energy to the
soft nanomagnet's potential energy. Although the generated strain is biaxial, we will approximate
it as uniaxial strain to somewhat compensate for the fact that not 100\% of the strain
in the PZT will be transferred to the soft nanomagnet. With this assumption, the stress anisotropy energy is written as
\begin{equation}
E_{str}(t) = - \frac{3}{2}\lambda_{s} \epsilon(t) Y \Omega {\cos}^2{\theta^{\prime}(t)},
\end{equation}
where $\lambda_s$ is the magnetostriction coefficient of the soft
nanomagnet, $Y$ is the Young's modulus, and $\epsilon(t)$
is the strain generated by the applied voltage $V_P(t)$ at the instant of time $t$.

The total potential energy of a {\it stressed} nanomagnet at any instant of time $t$ is therefore
\begin{equation}
E(t) = E \left ( \theta^{\prime}(t), \phi^{\prime}(t) \right ) = E_{sh}(t) + E_{m}(t) + E_{str}(t).
\end{equation}

We follow the standard procedure to derive the time evolution of the polar and azimuthal angles of the magnetization vector
of the soft nanomagnet in the rotated coordinate frame under the actions of the torques due
to shape anisotropy, stress anisotropy and magnetic field by solving the Landau-Lifshitz-Gilbert (LLG) equation:
\begin{eqnarray}
\frac{d\mathbf{m}(t)}{dt} & - &  \alpha \left[ \mathbf{m}(t) \times  \frac{d\mathbf{m}(t)}{dt} \right]
                           =  \frac{-|\gamma|}{\mu_0 M_s \Omega}  \mathbf{\tau_{ss}}(t),
\end{eqnarray}
where $\alpha$ is the Gilbert damping coefficient which depends on the soft nanomagnet's material, $\gamma$ is the gyromagnetic ratio
(a universal constant) and $\tau_{ss}$ is the total torque acting on the magnetization vector and is given by
\begin{eqnarray}
\mathbf{\tau_{ss}}(t)  & = &  - \mathbf{m}(t) \times \left(\frac{\partial E(t)}{\partial \theta^{\prime}(t)} \hat{\boldsymbol{\theta}} + \frac{1}{\sin \theta^{\prime}(t)}
\frac{\partial E(t)}{\partial \phi^{\prime}(t)} \hat{\boldsymbol{\phi}}\right) \nonumber\\
                     & = & \{E_{\phi 1}(t) \sin\theta^{\prime}(t) + E_{\phi 2}(t) \cos\theta^{\prime}(t) \nonumber\\
                     & - & \frac{1}{\sqrt{2}} M_s \Omega B \cos\phi^{\prime}(t)\}\hat{\boldsymbol{\theta}} \nonumber\\
                     & - & \{E_{s1} (t) \sin2\theta^{\prime}(t) + 2E_{s2} (t) \cos2\theta^{\prime}(t) \nonumber\\
                    & - & \frac{1}{\sqrt{2}} M_s \Omega B ( \sin\phi^{\prime}(t) \cos\theta^{\prime}(t) +  \sin\theta^{\prime}(t))  \nonumber\\
                     & + & (3/2) \lambda_s \epsilon(t) Y \Omega \sin2\theta^{\prime}(t) \}\hat{\boldsymbol{\phi}},
\end{eqnarray}
where $\mathbf{m} (t)$ is the normalized magnetization vector, quantities with carets are unit vectors in the original frame of reference, and
\begin{eqnarray}
E_{\phi 1}(t) & = & \frac{\mu_{0}}{4} {M}^2_{s} \Omega \{ \left( N_{d-yy}   + N_{d-zz}  \right) \sin 2\phi^{\prime}(t)
               -  2 N_{d-xx}\sin 2\phi^{\prime}(t) \} \nonumber \\
E_{\phi 2}(t) & = & \frac{\mu_{0}}{2} {M}^2_{s} \Omega \left( N_{d-zz} - N_{d-yy} \right) \cos \phi^{\prime}(t). \nonumber
\end{eqnarray}

At non-zero temperatures, there is an additional torque acting on the magnetization vector owing to thermal noise.
The procedure for finding this
torque has been described in Ref. \cite{Roy2012} and is not repeated here.

Solution of the LLG equation [Equation (8)] yields the magnetization of the stressed soft nanomagnet as a function of time
and steady state is achieved when the magnetization no longer changes appreciably with time.
This yields the switching time (time elapsed before reaching steady-state) and the
energy dissipation for any given stress. The procedure for finding these quantities
in the presence of thermal fluctuations requires solving the stochastic
LLG equation and is described in Ref. \cite{Roy2012} and \cite{Scholz2001}.

We assume that the soft nanomagnet is made of Terfenol-D which has the
following material parameters: saturation magnetization
$M_s =$ 8 $\times$ 10$^5$ A/m, magnetostriction coefficient $\lambda_s =$ 60 $\times$ 10$^{-5}$ ,
Young's modulus $Y$ = 80 GPa and
Gilbert damping coefficient
 $\alpha=$ 0.1
\cite{Abbundi1977,Ried1998,Kellogg2008}.

We solve the Landau-Lifshitz-Gilbert equation [Equation (8)] at 0 K (and its stochastic version at 300 K)
to find the steady-state orientation of the magnetization vector of the soft nanomagnet
as a function of stress in the nanomagnet and therefore as a function of the sum-total voltage applied at the
electrode pairs. The stress is varied between -50 and +50 MPa.
By following the procedure in Ref. \cite{Lynch2013}, we compute that an electric field of 37.5 kV/m
is required to produced a stress of 1 MPa in the PZT, which we assume is
fully transferred to the soft magnet.  Therefore, assuming that stress is linearly proportional to the voltage,
the corresponding voltages for $\pm$50 MPa are $\pm$187.5 mV
if the PZT film's thickness is 100 nm.

At the critical stress value (or critical voltage), the soft layer's magnetization vector changes abruptly from
its initial stable orientation to the other causing the angle between the magnetizations of the soft and hard layer to
change from $\beta_L = 180^{\circ} - 86.1^{\circ} = 93.9^{\circ}$ to $\beta_H = 180^{\circ}$, or vice versa.
This will cause the S-MTJ resistance to change by a factor of $\left (1 + \eta_1 \eta_2 cos (93.9^{\circ}) \right )
/\left ( 1 - \eta_1 \eta_2 \right )$ $\sim$ 1.9 if we assume $\eta_1 = \eta_2 = 0.7$.

Figure 2 shows the ratio $R(V)/R_L$ as a function of voltage $V$ applied at the contact pads, where
$R(V)$ is the MTJ resistance at a voltage $V$ and $R_L$ is the MTJ resistance in the low-resistance state.
If the MTJ is initially in the high resistance state, a compressive stress (positive voltage)
is required to drive it to the low resistance state, whereas if it is initially in the low resistance
state, a tensile stress (negative voltage) is required to drive it to the high resistance state because
we have assumed the soft magnet material to be Terfenol-D which has positive magnetostriction. Thus,
the critical voltage to switch from high-to-low resistance is +93.15 mV and the critical voltage to switch
from low-to-high resistance is -93.15 mV, which produces the appearance of a `hysteresis' in the characteristic
in Fig. 2.
This hysteretic behavior indicates that the device can also be used as a non-volatile memory.
Note that the transitions are not completely abrupt even at 0 K temperature
because sub-critical stress that is slightly lower than the critical stress
can cause some rotation of the soft magnet's magnetization vector and hence change the MTJ resistance
perceptibly. This is the reason for the `rounded corners'.

\begin{figure}[!ht]%
 \centering
  \includegraphics[width=3.4in]{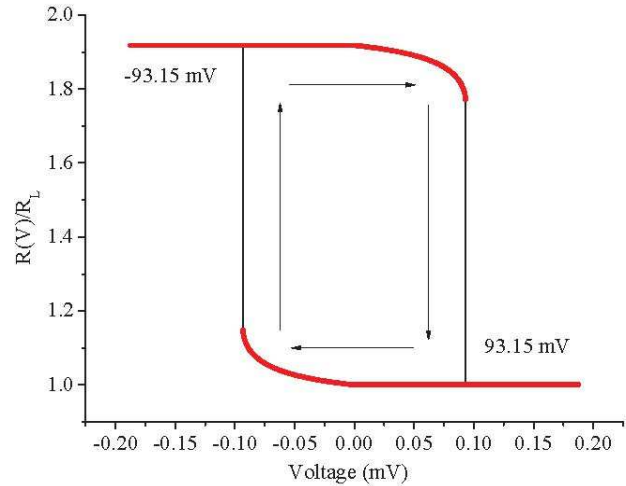}
 \caption[Caption of figure] {The transfer function (or firing behavior) for a straintronic
 spin neuron at 0 K. When the voltage appearing at node $P$ (in Figure 1) due to weighted inputs and
 bias is $V$, the
 resistance of the MTJ is $R(V)$. The low resistance is $R_L$. }
  \label{fig:MTJ_SN_0K_voltage}
\end{figure}

The energy dissipated during a firing event is the sum of the $CV^2$ dissipation associated with charging
the capacitance of the electrodes A and A$^{\prime}$,
and the internal dissipation in the soft magnet due to Gilbert damping. We neglect any dissipation
due to the current sources assuming that the currents are small and the current sources are turned on
only when the inputs arrive. Here,
$C$ is the capacitance of the electrodes and $V$ is the voltage at the electrodes. The capacitance
is determined from the areas of the electrodes, the dielectric constant of PZT and the PZT film
thickness. It is found to be 0.88fF.
Therefore, the capacitance charging dissipation will be $2 \times (1/2)CV^2$
= 0.09315$^2$ $\times$ 0.88fF = 7.63 aJ (there are two electrodes A and A$^{\prime}$). The dissipation due to Gilbert damping at 0 K was found
to be 1.2 aJ from the solution of the LLG equation as described in Ref. \cite{Roy2012} and \cite{Lyutyy2015}. Therefore,
the total
dissipation during a firing event
is 8.83 aJ. This is $\sim$450 times lower than that
reported (4 fJ) in Ref. \cite{KaushikRoy2013} for current-driven spin neurons when the
input voltage level was 100 mV and
$\sim$45 times lower than that (0.4 fJ) when the input voltage level was 10 mV (our input voltage level is $\sim$ 93 mV).
Note that the synapse resistances $r_1 \cdot \cdot \cdot r_N$ can be set arbitrarily
high and hence the dissipation in them can be neglected.
The switching (firing) delay is found to be 1.37 ns at the threshold voltage level.

\section{Current-driven Spin Neuron Based on Spin Transfer Torque (STT) switched MTJ}

In this section, we discuss a spin-neuron implemented with the same type of
MTJ as above, except this time the soft magnet is not switched
with strain, but with a spin polarized current delivering a spin transfer torque (STT).
The inputs are therefore not voltages, but currents.
We choose the material to be CoFeB with in-plane anisotropy which has a Gilbert damping constant of $\sim$0.004 \cite{Iihama2014}, much lower
than that of Terfenol-D. This material has in-plane anisotropy if the magnet thickness exceeds a few nm \cite{Ikeda2010}.
STT-driven spin neurons built with magnets exhibiting perpendicular anisotropy have been studied
in Refs. \cite{KaushikRoy2013} and \cite{KaushikRoy2014}.

Note that we chose two different materials -- Terfenol-D for the straintronic neuron and CoFeB for the current driven neuron --
because we wish to optimize both for minimal energy dissipation and then make a fair comparison. Terfenol-D has a very large
magnetostriction coefficient and is hence beneficial for a straintronic neuron. A current driven neuron does
not benefit from large magnetostriction, but benefits from small Gilbert damping, which is why we chose CoFeB for it.

\begin{figure}[!ht]%
 \centering
  \includegraphics[width=3.4in]{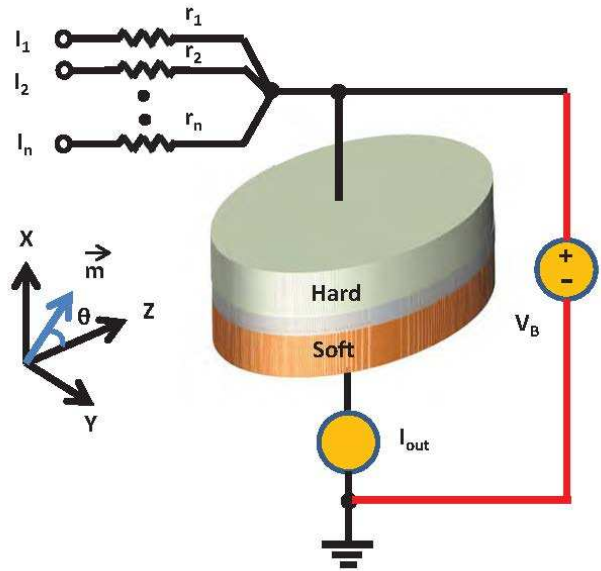}
 \caption[Caption of figure]{{\bf Schematic of a current-driven spin-neuron}}
  \label{fig:STT_SN}
\end{figure}

Figure 3 shows the schematic of a STT-based current-driven spin neuron consisting of an MTJ stack that is not skewed.
Both magnets are elliptical in shape and each has two stable magnetization orientations
along the major axis of the ellipse. Their major axes are collinear.

The hard magnet is permanently magnetized along one of its stable
orientations. The soft magnet's magnetization can be either parallel (low-resistance state) or anti-parallel
(high-resistance state) to that of the hard magnet. The high-to-low resistance ratio in this case is
$\left (1 + \eta_1 \eta_2 \right )/\left ( 1 - \eta_1 \eta_2 \right )$ = 2.9, assuming once again that
$\eta_1 = \eta_2 = 0.7$ \cite{Salis2005}. This corresponds to a tunneling magnetoresistance ratio,
or TMR, of $\sim$200\%.

A negative potential applied between the hard and soft layers will align the magnetization of the
soft layer parallel to that of the hard layer (low MTJ resistance), while a positive
potential will make it anti-parallel (high MTJ resistance).
These potentials cause a spin polarized current to be injected into or extracted from the soft layer which brings about
the magnetization switching by exerting a spin-transfer-torque
on the magnetization vector
\cite{Slonczewski1996,Ralph2008,Kubota2008,Rowlands2011}. For consistency, the soft layer's dimensions
are chosen such that the in-plane shape anisotropy energy barrier is 71.7 kT, close to
that of the soft magnet in the straintronic spin neuron. This is ensured by choosing the major axis $a$ = 50 nm,  minor axis $b$
 = 32 nm and thickness $d$ = 5 nm \cite{Beleggia2005}.

In the absence of  spin polarized current, the soft nanomagnet has only shape anisotropy energy and its
potential energy at any instant of time $t$ is given by,
\begin{eqnarray}
&E_{shape}(t) = E_{ss}(t) {\sin}^2{\theta(t)} + \frac{\mu_{0}}{2} \Omega {M}^2_{s} N_{d-zz} \nonumber\\
E_{ss}(t) &= \left( \frac{\mu_{0}}{2} \right) \Omega {M}^2_{s} \{ N_{d-xx}{\cos}^2{\phi(t)} + N_{d-yy}{\sin}^2{\phi(t)} - N_{d-zz} \},
\end{eqnarray}
where $\theta(t)$ and $\phi(t)$ are again, respectively, the instantaneous polar and azimuthal angles of the magnetization
vector, and $M_s$ is the saturation magnetization which is  $\sim$10.4 $\times$ 10$^5$ A/m \cite{Hayakawa2005} for amorphous CoFeB.

The torque on the magnetization vector at any time $t$ due to shape anisotropy can be expressed as
\begin{eqnarray}
\mathbf{\tau_{ss}}(t) &=& - \mathbf{m}(t) \times \left(\frac{\partial E(t)}{\partial \theta(t)} \hat{\boldsymbol{\theta}}
 + \frac{1}{\sin \theta(t)}
\frac{\partial E(t)}{\partial \phi(t)} \hat{\boldsymbol{\phi}}\right) \nonumber\\
  &=& E_{\phi s}(t) \sin\theta(t) \hat{\boldsymbol{\theta}} - E_{ss} (t) \sin2\theta(t) \hat{\boldsymbol{\phi}},
\end{eqnarray}
where $\mathbf{m} (t)$ is the normalized magnetization vector and\\
$E_{\phi s}(t) \!= \! \{\frac{\mu_{0}}{2} {M}^2_{s} \left(N_{d-yy} - N_{d-xx}\right)\} \Omega \sin2\phi(t).$ 

Passage of the spin-polarized current $I_s$ through the nanomagnet generates a spin transfer torque (STT) on the
magnetization vector given by \cite{Salahuddin2008}
\begin{eqnarray}
\mathbf{\tau_{sst}}(t)
&=& s [{\tt b} \sin(\zeta - \theta(t)) \hat{\boldsymbol{\phi}} - {\tt c} \sin(\zeta - \theta(t)) \hat{\boldsymbol{\theta}}],
\end{eqnarray}
where $s = (\hbar/2e) \chi I_s$ is the spin angular deposition per unit time, $\chi$ is the spin polarization of the current,
{\tt b} and {\tt c} are coefficients of the out-of-plane and
in-plane components of the spin-transfer-torque. We assume $\chi$ = 70\% (assuming 70\% spin injection
efficiency), and {\tt b} and {\tt c} are 0.3 and 1, respectively.
The current is passed perpendicular to the
plane of the magnet as shown in Figure 3. The quantity $\zeta$ is the angle subtended by the direction of
spin polarization with the z-axis and it is either 0$^{\circ}$ or 180$^{\circ}$.

The Landau-Lifshitz-Gilbert (LLG) equation and its stochastic version are solved again to extract the
STT-induced magnetization switching behavior of the soft nanomagnet in the absence (0 K) and presence (300 K) of thermal noise.
\begin{eqnarray}
\frac{d\mathbf{m}(t)}{dt} & - & \alpha \left( \mathbf{m}(t) \times  \frac{d\mathbf{m}(t)}{dt} \right) = \frac{-|\gamma|}{\mu_0 M_s \Omega} \left ( \mathbf{\tau_{ss}}(t) +\mathbf{\tau_{sst}}(t) \right )
\end{eqnarray}

\begin{figure}[!ht]%
 \centering
  \includegraphics[width=3.4in]{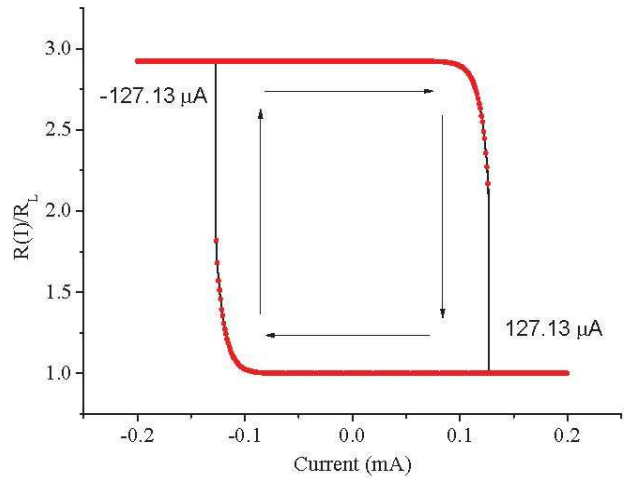}
 \caption[Caption of figure] {The transfer characteristic (or firing behavior) for a STT-based
 current-driven spin neuron at 0 K. $R(I)$ is the  resistance of the MTJ when the total current injected into the
 soft layer (due to weighted inputs and bias) is $I$ and $R_L$ is again the low resistance of the MTJ.}
\end{figure}

First, we determine the current required to switch the soft layer (from the LLG simulations) and find it to be
127.13 $\mu$A (the corresponding current density is 10.11 MA/cm$^2$) for a current pulse of duration 10 ns.
Ref. \cite{Amiri2011} also assumed a current pulse duration of 10 ns and found the switching current density to be
4 MA/cm$^2$ for a barrier height of 42 kT in perpendicular magnetic anisotropy (our barrier height is 71.7 kT).
Ref. \cite{Tomasello2014} considered a CoFeB soft nanomagnet with in-plane anisotropy barrier
of $\sim$57 kT (calculated from the reported dimensions of 170 nm $\times$ 60 nm $\times$ 1.4 nm using Ref.
\cite{Beleggia2005}) and
found
the switching current density to be $\sim$ 5.5 MA/cm$^2$ for a switching time of 10 ns.

Figure 4 shows the 0 K temperature transfer characteristic (or firing behavior) of the
MTJ when the total input current $I$ has been varied between -200 $\mu$A and +200 $\mu$A. Note that
 rounded corners are more ``square'' here because a subthreshold current (unlike subthreshold voltage) hardly rotates the magnetization of the
soft magnet and hence does not alter the MTJ resistance perceptibly.

If we consider a typical resistance-area product (RA) of 2.1 $\Omega$-$\mu$m$^2$ for the MTJ  in the
high resistance state \cite{Isogami2008} and a TMR ratio of 200\%,
then the high-state MTJ resistance for our chosen dimensions becomes 1671 ohms and the low-state
MTJ resistance 557 ohms. Energy dissipation $E_R$ due to the current passing through MTJ's tri-layered structure
is given by \cite{Carpentieri2012}
\begin{equation}
E_{R} = \int_0^{\tau} dt I^2 \left[ R_P + (R_{AP} - R_P) \left(\frac{1-\cos{\theta(t)}}{2}\right)\right],
\end{equation}
where  $R_P$ and $R_{AP}$ are the MTJ resistances in the parallel (low) and anti-parallel (high) state, and
$\theta(t)$ is the angle between the magnetization states of the soft layer and the hard layer
at time $t$. This dissipation turns out to be $\sim$0.26 pJ.
The LLG simulation showed that the neuron takes 14 ns to switch.
The dissipation due to Gilbert damping in the soft magnet
is a mere $\sim$0.48 aJ, which is negligible.

The total dissipation reported in
Ref. \cite{KaushikRoy2013} for current-driven spin neurons that use magnets with perpendicular
anisotropy is 0.4 fJ. We used magnets with in-plane anisotropy. There are at least two reasons why Ref. \cite{KaushikRoy2013}
could have reported a 650 times lower dissipation compared to what we found (0.26 pJ). First, their
critical current density was 4 MA/cm$^2$ which is 2.5 times less than ours. Presumably, the lower current
density is due to the fact that the energy barrier between stable magnetization states in their
device might have been only $\sim$40 kT which is what we estimate following the procedure in Refs. \cite{Lee2011} and
\cite{Ikeda2010}. Ours was 71.7 kT. The critical current density scales with the energy barrier height; for example, Ref. \cite{Ikeda2007}
reported a critical current density of 8.7 MA/cm$^2$ for a barrier height of 67 kT.
Additionally, in Ref. \cite{KaushikRoy2013}, the  soft layer thickness was only 2 nm to maintain perpendicular anisotropy,
whereas our thickness was 5 nm and the critical current density increases with the soft layer thickness \cite{Hayakawa2005}.
These two factors increased the current density in our case and caused a higher dissipation.
Second, and more importantly,
Ref. \cite{KaushikRoy2013} utilized the spin Hall effect to inject/extract spin polarized current from the magnets
which would allow passing the charge current parallel to the heterointerface between the spacer and the magnets.
This would allow the current to avoid going through the highly resistive spacer and decrease the resistance-area product of
the MTJ considerably compared to ours. These two factors might be the cause for the 650 times lower dissipation figure reported in
Ref. \cite{KaushikRoy2013} compared to what we find. Even then, the lower dissipation reported in Ref. \cite{KaushikRoy2013}
is still 45 times higher than that encountered in the straintronic spin neuron. If we carry out the comparison between similar designs
(in-plane anisotropy magnets, similar energy barrier heights to maintain similar resilience to thermally induced random
firing), then the difference is even more stark; the straintronic spin neuron is 29445 times more energy-efficient and yet 10
times faster. The current-driven spin neuron however has however one small advantage; it has a 50\%
higher on/off ratio of the MTJ resistance because the angular separation between the two stable orientations of the
soft magnet in the MTJ
is $\sim180^{\circ}$ for the current-driven spin neuron and $\sim90^{\circ}$ for the straintronic spin neuron.

\begin{figure}[!ht]%
 \centering
  \includegraphics[width=3.4in]{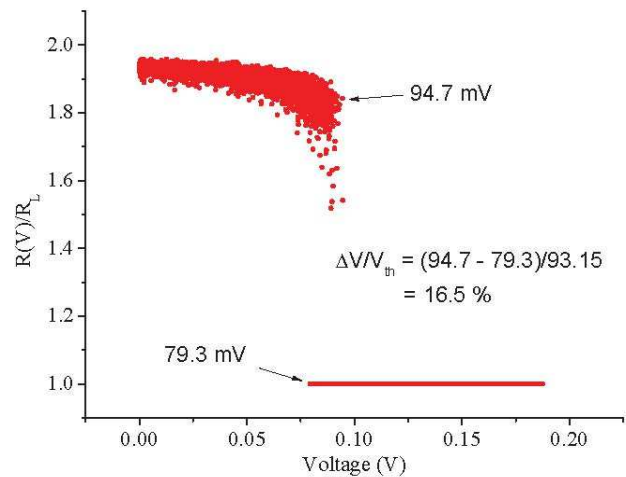}
 \caption[Caption of figure] {Transfer characteristic (or firing behavior) for the straintronic spin neuron at room temperature in the presence of thermal
 noise. Results are shown for positive threshold only since no additional information
 can be gleaned from the negative threshold segment. Since the simulation is terminated immediately upon completion of
 firing, no fluctuations in the transfer characteristic
 are visible in the low-resistance state.}
  \label{fig:MTJ_SN_300K_voltage}
\end{figure}

\begin{figure}[!ht]%
 \centering
  \includegraphics[width=3.4in]{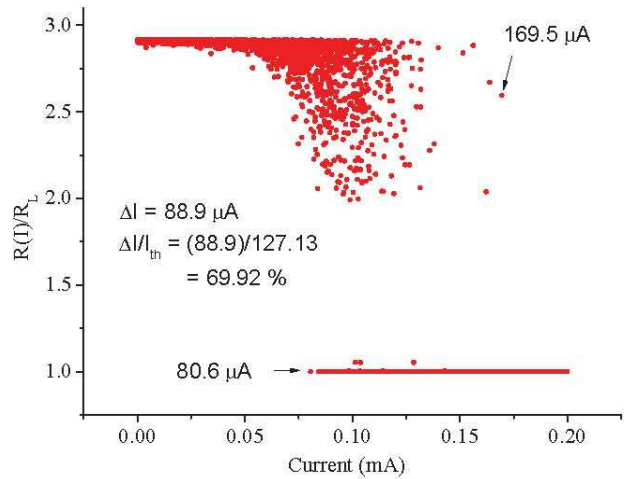}
 \caption[Caption of figure]{Transfer characteristic (or firing behavior) for the current-driven spin neuron at room temperature in the presence of thermal
 noise.}
  \label{fig:STT_SN_300K_voltage}
\end{figure}

\section{Room-temperature firing behavior of spin neurons}
The room temperature firing characteristics are found by solving the stochastic LLG equation
which will generate a distribution of characteristics since each one is slightly different in the presence of
thermal noise. Each characteristic will show slightly different switching threshold and slightly different switching
delay.

Figure 5 shows the $R(V)/R_L$ versus $V$ characteristic of the straintronic spin neuron at 300 K when switching
from high- to low-resistance state (the characteristic for switching from low- to high-resistance
state will be qualitatively similar and hence not shown). As many as 10,000 switching trajectories were
simulated from the stochastic LLG equation and make up this plot. There average switching time
was 0.61 ns when the voltage appearing at node $P$ in Figure 1 was above the 0 K threshold voltage of 93.15 mV.
Clearly, the threshold has been significantly
broadened at room temperature, indicating that there is a significant probability of
pre-mature firing (firing before reaching the threshold defined at 0K) because of random thermal torque,
as well as failure to fire at or slightly beyond the threshold for the same reason.
The likelihood of false firing  is a matter of concern which calls for further investigation of thermal degradation.

One measure of the threshold broadening is the ratio $\Delta V$/$V_{th}$, where $V_{th}$ is the average
voltage at which the neuron fires and $\Delta V$ is the standard deviation.
For the straintronic spin neuron studied, this quantity turns out to be 16.5\%. Of course this can be
reduced by increasing $V_{th}$ (by making the soft nanomagnet more shape-anisotropic to increase the
in-plane shape anisotropy energy), but this will also increase the energy dissipation which varies
roughly as $V_{th}^2$ (because the dissipation is dominated by the $CV^2$ loss). Let us say that 1\% broadening is acceptable. Then we will have to increase
$V_{th}$ 16.5-fold (assuming that it does not change $\Delta V$), resulting in an energy dissipation of 8.83aJ $\times$ 16.5$^2$ = 2.4 fJ. This is still much less
than what is encountered in CMOS-base implementations (0.7 pJ reported in Ref. \cite{KaushikRoy2013}).

The same thermal broadening is not only present in a current-driven spin neuron, but it is much worse there.
Figure 6 shows $R(I)/R_L$ versus $I$
for the current-driven spin neuron where $I$ is the driving current. Here, 5,000 switching trajectories were simulated to
produce this plot because simulation of 10,000 trajectories became computationally prohibitive (because of the
longer simulation time). The average
switching time was now 7.26 ns above the 0 K threshold current of 127.13 $\mu$A.
The quantity $\Delta I$/$I_{th}$ turns out to be 69.92\% for the simulated neuron. Had we been able to simulate
more trajectories, we might have found the broadening to be even worse.
Once again, the relative broadening can be reduced by increasing $I_{th}$ (by increasing the shape anisotropy of the
soft nanomagnet)
but of course at the cost of increasing dissipation since the latter
varies roughly as $I_{th}^2$ (because the dissipation is dominated by the $I^2R$ loss). Once again, if 1\% broadening is acceptable, then we will need to increase the
threshold 69-fold (assuming this does not change $\Delta I$), resulting in an energy dissipation of 0.26pJ $\times$ 69$^2$ = 1.23 nJ. This would make it 1768 times
worse than CMOS-based neurons and therefore the viability of STT-based spin neurons at room temperature is questionable.
Even if better design (use of spin Hall effect, perpendicular anisotropy magnets, etc.) can reduce the energy dissipation by
two orders of magnitude, it will still provide little advantage over CMOS implementations. In contrast, the straintronic
spin neuron is still $\sim$290 times more energy-efficient than its CMOS-based counterpart at room temperature.

\section{Conclusion}

In conclusion, we have proposed and analyzed a straintronic spin-neuron that is orders of magnitude more energy-efficient
than the
usual current-driven spin neuron and also faster. The primary obstacle they both face is the significant broadening of the
firing threshold at room temperature. For the current-driven spin neuron, it may not be possible to mitigate this problem
without increasing the energy dissipation to the point where it is no longer superior to CMOS-based implementations.
Fortunately, for the straintronic spin neuron, it may still be possible to mitigate this problem while retaining a significant
energy advantage over CMOS.

\section{Appendix}
The broadening in Fig. 6 for current-driven spin neuron due to thermal noise at room temperature can be reduced
if we choose a different soft material with larger Gilbert damping coefficient ($\alpha$). Since the switching current density
 is approximately linearly proportional to $\alpha$, the
energy dissipation will increase quadratically. In order to illustrate this, we chose a hypothetical material
which is  identical to CoFeB in all respects, except its Gilbert damping coefficient $\alpha$ is 0.1
and carried out the stochastic LLG simulations. The switching current
turned out to be 1.44 mA for a current pulse duration of 10 ns, but the broadening ($\Delta$I/I$_{th}$) was reduced to 11.25\%. However, the extremely large
switching current results in energy dissipation of  23.8 pJ which is clearly prohibitive. This case study
shows us that although thermal degradation may be countered with increased energy dissipation,
the price to be paid in energy may be prohibitive.

\noindent {\bf Programmable synapses}: In this paper, our synapses have fixed weights because they are implemented
with passive resistors. {\it Programmable} synapses are required for more versatile renditions of
neural computing such as Spike Time Dependent Plasticity (STDP) models of neural
networks \cite{Bi1998, Favero2014, Aoki2006, Sengupta2015} that are popular for Hebbian learning \cite{Hebb1949}. For STDP models, the synapse weight
should be programmed by a spike signal, e.g. it should be a function of a time-integrated spike current.
The time integrated spike current is a charge that could be
stored in a capacitor which then applies a voltage across a piezoelectric layer that generates strain and
rotates the magnetization of a magnetostrictive magnet that is elastically coupled to the piezoelectric layer.
The magnetostrictive magnet could be the soft layer of a magneto-tunneling junction whose resistance is thus programmed
by the time integrated spike signal, resulting in the appropriate programmable synapse. This is a different topic
and would be treated elsewhere.

This work was supported by the US National Science Foundation under grants ECCS-1124714 and CCF-1216614. J. A. would also like to acknowledge the NSF CAREER grant CCF-1253370.


\section*{References}
\begin{thebibliography}{10}
\bibitem{Chua1988} L. Chua and L. Yang, "Cellular neural networks: applications," {\it IEEE Trans. Circuits Syst.} \textbf{35}, 1273 (1988).
%
\bibitem{Ueda2011} M. Ueda, Y. Kaneko, Y. Nishitani, and E. Fujii, "A neural network circuit using persistent interfacial conducting heterostructures," \textit{J. Appl. Phys.} \textbf{110}, 086104 (2011).
%
\bibitem{Simard2003} P. Simard, D. Steinkraus, and J. C. Platt, "Best practices for convolutional neural networks applied to visual document analysis," in proceedings of the \textit{International Conference on Document Analysis and Recognition (ICDAR)} \textbf{3}, 958 (2003).
%
\bibitem{George2009} D. George and J. Hawkins, "Towards a mathematical theory of cortical micro-circuits," \textit{PLOS Comput. Biol.} \textbf{5}, e1000532 (2009).
%
\bibitem{Lippmann1987} R. Lippmann, "An introduction to computing with neural nets," \textit{IEEE ASSP Mag.} \textbf{2}, 4 (1987).
%
\bibitem{Datta2012} S. Datta, S. Salahuddin, and B. Behin-Aein, "Non-volatile spin switch for boolean and non-boolean logic," \textit{Appl. Phys. Lett.} \textbf{101}, 252411 (2012).
%
\bibitem{KaushikRoy2012} M. Sharad, C. Augustine, G. Panagopoulos, and K. Roy, "Spin-based neuron model with domain-wall magnets as synapse," \textit{IEEE Trans. Nanotech.} \textbf{11}, 843 (2012).
%
\bibitem{KaushikRoy2013} M. Sharad, D. Fan, and K. Roy, "Spin-neurons: A possible path to energy-efficient neuromorphic computers," \textit{J. Appl. Phys.} \textbf{114}, 234906 (2013).
%
\bibitem{Liu2012} L. Liu, C.-F. Pai, Y. Li, H. W. Tseng, D. C. Ralph, and R. A. Buhrman, "Spin-torque switching with the giant spin hall effect of tantalum," \textit{Science} \textbf{336}, 555 (2012).
%
\bibitem{Ramesh2007} F. Zavaliche, T. Zhao, H. Zheng, F. Straub, M. P. Cruz, P.-L. Yang, D. Hao, and R. Ramesh, "Electrically assisted magnetic recording in multiferroic nanostructures," \textit{Nano Lett.} \textbf{7}, 1586 (2007).
%
\bibitem{Brintlinger2010} T. Brintlinger, S.-H. Lim, K. H. Baloch, P. Alexander, Y. Qi, J. Barry, J. Melngailis, L. Salamanca-Riba, I. Takeuchi, and J. Cumings, In situ observation of reversible nanomagnetic switching induced by electric fields, \textit{Nano Lett.} \textbf{10}, 1219 (2010).
%
\bibitem{Atulasimha2010} J. Atulasimha and S. Bandyopadhyay, Bennett clocking of nanomagnetic logic using multiferroic single domain nanomagnets, \textit{Appl. Phys. Lett.} \textbf{97}, 173105 (2010).
%
\bibitem{Buzzi2013} M. Buzzi, R. Chopdekar, J. L. Hockel, A. Bur, T. Wu, N. Pilet, P. Warnicke, G. P. Carman, L. J. Heyderman, and F. Nolting, Single domain spin manipulation by electric fields in strain coupled artificial multiferroic nanostructures, \textit{Phys. Rev. Lett.} \textbf{111}, 027204 (2013).
%
\bibitem{DSouzaArxiv} N. DSouza, M. Salehi-Fashami, S. Bandyopadhyay, and J. Atulasimha, Strain induced clocking of nanomagnets for ultra low power boolean logic, \textit{arXiv:1404.2980 [cond-mat.mes-hall]} (2014).
%
\bibitem{Tiercelin2011} N. Tiercelin, Y. Dusch, V. Preobrazhensky, and P. Pernod, Magnetoelectric memory using orthogonal magnetization states and magnetoelastic switching, \textit{J. Appl. Phys.} \textbf{109}, 07D726 (2011).
%
\bibitem{Pertsev2009} N. A. Pertsev and H. Kohlstedt, Magnetic tunnel junction on a ferroelectric substrate, \textit{Appl. Phys. Lett.} \textbf{95}, 163503 (2009).
%
\bibitem{Roy2013} K. Roy, S. Bandyopadhyay, and J. Atulasimha, Binary switching in a symmetric potential landscape, \textit{Sci. Rep.} \textbf{03}, 3038 (2013).
%
\bibitem{Biswas2014a} A. K. Biswas, S. Bandyopadhyay, and J. Atulasimha, Energy-efficient magnetoelastic non-volatile memory, \textit{Appl. Phys. Lett.} \textbf{104}, 232403 (2014).
%
\bibitem{Biswas2014b} A. K. Biswas, S. Bandyopadhyay, and J. Atulasimha, Complete magnetization reversal in a magnetostrictive nanomagnet with voltage generated stress: A reliable energy-efficient non-volatile magnetoelastic memory, \textit{Appl. Phys. Lett.} \textbf{105}, 072408 (2014).
%
\bibitem{Wang2014} J. J. Wang, J. M. Hu, J.Ma, J. X. Zhang, L. Q. Chen, and C. W. Nan, Full 180$^{\circ}$ magnetization reversal with electric fields, \textit{Sci. Rep.} \textbf{04}, 7507 (2014).
%
\bibitem{Mohammed2011} M. S. Fashami, K. Roy, J. Atulasimha, and S. Bandyopadhyay, "Magnetization dynamics, bennett clocking and associated energy dissipation in multiferroic logic," \textit{Nanotechnology} \textbf{22}, 155201 (2011).
%
\bibitem{Mohammed2012} M. S. Fashami, J. Atulasimha, and S. Bandyopadhyay, "Magnetization dynamics, throughput and energy dissipation in a universal multiferroic nanomagnetic logic gate with fan-in and fan-out," \textit{Nanotechnology} \textbf{23}, 105201 (2012).
%
\bibitem{Biswas2014c} A. K. Biswas, J. Atulasimha, and S. Bandyopadhyay, "An error-resilient non-volatile magneto-elastic universal logic gate with ultralow energy-delay product," \textit{Sci. Rep.} \textbf{4}, 7553 (2014).
%
\bibitem{Salis2005} G. Salis, R. Wang, X. Jiang, R. M. Shelby, S. S. P. Parkin, S. R. Bank, and J. S. Harris, "Temperature independence of the spin-injection efficiency of a MgO-based tunnel spin injector," \textit{Appl. Phys. Lett.} \textbf{87}, 262503 (2005).
%
\bibitem{Ikeda2008} S. Ikeda, J. Hayakawa, Y. Ashizawa, Y. M. Lee, K. Miura, H. Hasegawa, M. Tsunoda, F. Matsukura, and H. Ohno, "Tunnel magnetoresistance of 604\% at 300 K by suppression of Ta diffusion in CoFeB/MgO/CoFeB pseudo-spin-valves annealed at high temperature," \textit{Appl. Phys. Lett.} \textbf{93}, 082508 (2008).
%
\bibitem{Lynch2013} J. Cui, J. L. Hockel, P. K. Nordeen, D. M. Pisani, C. y. Liang, G. P. Carman, and C. S. Lynch, "A method to control magnetism in individual strain-mediated magnetoelectric islands," \textit{Appl. Phys. Lett.} \textbf{103}, 232905 (2013).
%
\bibitem{Cowburn1999} R. P. Cowburn, D. K. Koltsov, A. O. Adeyeye, M. E. Welland, and D. M.Tricker, "Single-domain circular nanomagnets," \textit{Phys. Rev. Lett.} \textbf{83}, 1042 (1999).
%
\bibitem{Beleggia2005} M. Beleggia, M. D. Graef, Y. T. Millev, D. A. Goode, and G. Rowlands, "Demagnetization factors for elliptic cylinders," \textit{J. Phys. D: Appl. Phys.} \textbf{38}, 3333 (2005).
%
\bibitem{Roy2012} K. Roy, S. Bandyopadhyay, and J. Atulasimha, "Energy dissipation and switching delay in stress-induced switching of multiferroic nanomagnets in the presence of thermal fluctuations," \textit{J. Appl. Phys.} \textbf{112}, 023914 (2012).
%
\bibitem{Scholz2001} W. Scholz, T. Schrefl, and J. Fidler, "Micromagnetic simulation of thermally activated switching in fine particles," \textit{J. Magn. Magn. Mater.} \textbf{233}, 296 (2001).
%
\bibitem{Abbundi1977} R. Abbundi and A. E. Clark, "Anomalous thermal expansion and manetostriction of single crystal Tb$_{0.27}$Dy$_{0.73}$Fe$_{2}$," \textit{IEEE Trans. Magn.} \textbf{13}, 1519 (1977).
%
\bibitem{Ried1998} K. Ried, M. Schnell, F. Schatz, M. Hirscher, B. Ludescher, W. Sigle, and H. Kronm$\ddot{u}$ller, "Crystallization behaviour and magnetic properties of magnetostrictive TbDyFe films," \textit{Phys. Status Solidi A} \textbf{167}, 195 (1998).
%
\bibitem{Kellogg2008} R. Kellogg and A. Flatau, "Experimental investigation of Terfenol-D's elastic modulus," \textit{J. Intell. Mater. Syst. Struct.} \textbf{19}, 583 (2008).
%
\bibitem{Lyutyy2015} T. V. Lyutyy, S. I. Denisov, A. Y. Peletskyi, and C. Binns, "Energy dissipation in single-domain ferromagnetic nanoparticles: Dynamical approach," \textit{arXiv:1502.04222v1[cond-mat.mes-hall]} (2015).
%
\bibitem{Iihama2014} S. Iihama, S. Mizukami, H. Naganuma, M. Oogane, Y. Ando, and T. Miyazaki, "Gilbert damping constants of Ta/CoFeB/MgO(Ta) thin films measured by optical detection of precessional magnetization dynamics," \textit{Phys. Rev. B} \textbf{89}, 174416 (2014).
%
\bibitem{Ikeda2010} S. Ikeda, K. Miura, H. Yamamoto, K. Mizunuma, H. D. Gan, M. Endo, S. Kanai, J. Hayakawa, F. Matsukura, and H. Ohno, "A perpendicular-anisotropy CoFeB-MgO magnetic tunnel junction," \textit{Nature Mater.} \textbf{9}, 721 (2010).
%
\bibitem{KaushikRoy2014} M. Sharad, D. Fan, K. Aitken, and K. Roy, "Energy-efficient non-boolean computing with spin neurons and resistive memory," \textit{IEEE Trans. Nanotech.} 13, 23 (2014).
%
\bibitem{Slonczewski1996} J. Slonczewski, "Current-driven excitation of magnetic multilayers," \textit{J. Magn. Magn. Mater.} \textbf{159}, L1 (1996).
%
\bibitem{Ralph2008} D. C. Ralph and M. D. Stiles, "Spin transfer torques," \textit{J. Magn. Magn. Mater.} \textbf{320}, 1190 (2008).
%
\bibitem{Kubota2008} H. Kubota, A. Fukushima, K. Yakushiji, T. Nagahama, S. Yuasa, K. Ando, H. Maehara, Y. Nagamine, K. Tsunekawa, D. D. Djayaprawira, N. Watanabe, and Y. Suzuki, "Quantitative measurement of voltage dependence of spin-transfer torque in MgO-based magnetic tunnel junctions," \textit{Nature Phys.} \textbf{4}, 37 (2008).
%
\bibitem{Rowlands2011} G. E. Rowlands, T. Rahman, J. A. Katine, J. Langer, A. Lyle, H. Zhao, J. G. Alzate, A. A. Kovalev, Y. Tserkovnyak, Z. M. Zeng, H. W. Jiang, K. Galatsis, Y. M. Huai, P. K. Amiri, K. L. Wang, I. N. Krivorotov, and J.-P. Wang, "Deep subnanosecond spin torque switching in magnetic tunnel junctions with combined in-plane and perpendicular polarizers," \textit{Appl. Phys. Lett.} \textbf{98}, 102509 (2011).
%
\bibitem{Hayakawa2005} J. Hayakawa, S. Ikeda, Y. M. Lee, R. Sasaki, T. Meguro, F. Matsukura, H. Takahashi, and H. Ohno, "Current-driven magnetization switching in CoFeB/MgO/CoFeB magnetic tunnel junctions," \textit{Jap. J. Appl. Phys.} \textbf{44}, L1267 (2005).
%
\bibitem{Salahuddin2008} S. Salahuddin, D. Datta, and S. Datta, "Spin transfer torque as a non-conservative pseudo-field," \textit{arXiv:0811.3472 [cond-mat.mes-hall]} (2008).
%
\bibitem{Amiri2011} P. K. Amiri, Z. M. Zeng, J. Langer, H. Zhao, G. Rowlands, Y.-J. Chen, I. N. Krivorotov, J.-P. Wang, H. W. Jiang, J. A. Katine, Y. Huai, K. Galatsis, and K. L. Wang, "Switching current reduction using perpendicular anisotropy in CoFeB/MgO magnetic tunnel junctions," \textit{Appl. Phys. Lett.} \textbf{98}, 112507 (2011).
%
\bibitem{Tomasello2014} R. Tomasello, V. Puliafito, B. Azzerboni, and G. Finocchio, "Switching properties in magnetic tunnel junctions with interfacial perpendicular anisotropy: Micromagnetic study," \textit{IEEE Trans. Magn.} \textbf{50}, 7100305 (2014).
%
\bibitem{Isogami2008} S. Isogami, M. Tsunoda, K. Komagaki, K. Sunaga, Y. Uehara, M. Sato, T. Miyajima, and M. Takahashi, "In situ heat treatment of ultrathin mgo layer for giant magnetoresistance ratio with low resistance area product in CoFeB/MgO/CoFeB magnetic tunnel junctions," \textit{Appl. Phys. Lett.} \textbf{93}, 192109 (2008).
%
\bibitem{Carpentieri2012} M. Carpentieri, M. Ricci, P. Burrascano, L. Torres, and G. Finocchio, "Wideband microwave signal to trigger fast switching processes in magnetic tunnel junctions," \textit{J. Appl. Phys.} \textbf{111}, 07C909 (2012).
%
\bibitem{Lee2011} K. Lee, J. J. Sapan, S. H. Kang, and E. E. Fullerton, "Perpendicular magnetization of CoFeB on single crystal MgO," \textit{J. Appl. Phys.} \textbf{109}, 123910 (2011).
%
\bibitem{Ikeda2007} S. Ikeda, J. Hayakawa, Y. M. Lee, F. Matsukura, H. Ohno, and T. Hanyu, "Magnetic tunnel junctions for spintronic memories and beyond," \textit{IEEE Trans. Elec. Dev.} \textbf{54}, 991 (2007).
%
\bibitem{Bi1998}
G-q Bi and M-m Poo, Synaptic modifications in cultured hippocampal neurons: Dependence on spike timing, synaptic strength,
 and postsynaptic cell type, \textit{J. Neuroscience} \textbf{18}, 10464 (1998).
%
\bibitem{Favero2014}
 M. Favero, A. Cangiano and G. Busetto, Hebb-based rules of neuro-plasticity: Are they ubiquitously important for the refinement
 of synaptic connections in development, \textit{The Neuroscientist} \textbf{20}, 8 (2014).
%
\bibitem{Aoki2006}
 T. Aoki and T. Aoyagi, A possible role of incoming spike synchrony in associative memeory model with STDP learning rule,
 \textit{Prog. Theor. Phys. Suppl.} \textbf{161}, 152 (2006).
%
\bibitem{Sengupta2015}
A. Sengupta, Z. Al Azim, X. Fong and K. Roy, Spin-orbit torque induced spike-timing dependent plasticity, \textit{Appl. Phys. Lett.} \textbf{106}, 093704 (2015).
%
\bibitem{Hebb1949}
D. Hebb, {\it The Organization of Behavior} (Wiley, New York, 1949).

\end{thebibliography}
\end{document}